*2024-2025 CRA Quadrennial Paper*

# Lessons for Cybersecurity from the American Public Health System

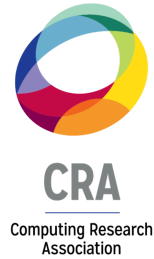
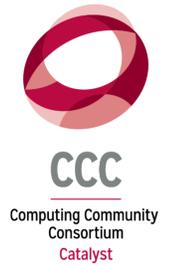

Adam Shostack (University of Washington), L. Jean Camp (Indiana University), Yi Ting Chua (University of Tulsa), Josiah Dykstra (Trail of Bits), Brian LaMacchia (FARCASTER Consulting Group), Daniel Lopresti (Lehigh University)

> **The United States needs national institutions and frameworks to systematically collect cybersecurity data, measure outcomes, and coordinate responses across government and private sectors, similar to how public health systems track and address disease outbreaks.**

## Public Health and Cybersecurity Public Health

Public health is a discipline focused on the health of populations. Public health and medicine complement each other, and their advances lead to measurable extensions of human life, such as nearly doubling population life expectancy during the 20th century. Public health allows for the comparison of alternative courses of treatment for best effectiveness and enables the allocation of limited resources to have the greatest possible impact on the largest at-risk population.

The advantage of taking a public health approach to disease is exemplified by one of its earliest examples. In 1854, while England — and particularly London — was suffering through an epidemic of cholera, physician John Snow theorized that the disease spread via water rather than air, as was assumed. To test his theory, Snow took a novel approach of mapping the locations of cholera deaths in the city and city water pumps. He noticed that deaths appeared to be disproportionately clustered around a particular water pump on Broad Street. When he removed the pump handle,incidences of cholera dropped considerably. Snow also performed a statistical analysis of cholera deaths among customers of two different water companies drawing from different parts of the Thames River — one that drew close to the city and one that drew further upstream, and therefore likely less polluted from city sewage. The population served by the upstream water company had 14 times fewer cholera deaths, further strengthening his hypothesis. It was a convincing demonstration of the value of a public health approach, combining medical knowledge and data with spatial and statistical data to point to an effective course of action.

The current state of cybersecurity outcome research is very similar to medical research before the establishment of the field of public health. Knowing that cholera deaths decreased after

John Snow removed a pump handle enabled evaluation of his theory that the disease was waterborne. If someone claimed to have an equivalent intervention for a problem in cybersecurity, how could we measure its effectiveness? This is not a pure "metrics" question but rather one that requires understanding changes in many systems.

In public health, three core categories are measured: populations, outcomes, and vectors. For example:

- Population: All the people in Mississippi.
- Outcome: Deaths or quality-adjusted years of life lost.
- Vectors: Bird flu and bubonic plague.

Defining analogous categories in cyberspace is much more complex. The cybersecurity community is, in a sense, working like early 19th century physicians, developing new cybersecurity defensive technologies without rigorous ways to measure their effectiveness at improving outcomes across large populations. Cybersecurity researchers and practitioners still struggle to measure and categorize the propagation of attack vectors across our increasingly interconnected cybersecurity infrastructure.

Indeed, there are a surprising number of things we still do not know, for example:

- Does running antivirus software reduce the incidence of successful attacks?
- How much would it cost to reduce the likelihood of a successful attack by half?
- What's the best outcome that can be obtained for a 10 percent budget increase?
- When should the transition to post-quantum cryptography occur?
- How many data breaches were there in 2024? (We know about a subset of those that impacted personal data, but we don't know how many of those were either undetected or concealed.)
- How many computers are in active use, or what portion run specific operating systems? (We have statistics about computers used to browse the web, but that excludes IoT devices such as baby monitors, smart doorbells, televisions, cars, or jet engines. At least, we hope none of those are browsing the web.)

It is difficult to know if American society is winning against cyber threats. Researchers do not know if we are more secure today than yesterday. And when researchers want to measure the impact of initiatives like the National Cybersecurity Strategy, the measures they need are missing.

What is needed is the adoption of a public health like approach to cybersecurity.

Some indicators that would prove Cybersecurity Public Health (CPH) investments are making a difference could include:

- A decline in successful attacks, according to a consistent data collection system.

- Publication of a cybersecurity equivalent of a morbidity and mortality weekly report would show that an institution has sufficient data gathering and analysis capabilities to detect new vectors or outcomes that deserve community attention and the credibility to deserve such attention.

- Increased device lifespans, reducing costs and e-waste.

- Updated standards to remove demonstrably ineffective techniques, like password rotation.

The Federal Government must play a role. Cybersecurity challenges transcend individual organizations or industries. The Federal Government is uniquely positioned to coordinate efforts on a national scale, as it does with public health initiatives. Establishing CPH as a discipline would require resources for research, infrastructure, and implementation. The Federal Government can allocate funds and resources at a level, and with a patience, that individual organizations cannot match, particularly to drive research, development, and institutional support. Furthermore, the government can facilitate the collection and sharing of cybersecurity data across public and private sectors, which is crucial for developing effective CPH strategies, similar to how the CDC collects and disseminates public health data.

# Recommendations

## Establish a Bureau of Cyber Public Health Statistics

The most important step the government could take to further CPH at this time would be to establish a Bureau of Cyber Public Health Statistics (similar to the recommendation in the Solarium Report), and charge it with measuring the cybersecurity health of American society, as the Centers for Disease Control and Prevention (CDC) does for health. It will identify gaps in existing knowledge and highlight the challenges they face, which may necessitate funding for science, advanced research projects, or other activities.

# Enable the Federal Government to Measure and Improve Cybersecurity Controls

Another critical step is to empower part of the government to measure outcomes, compare them to existing controls, and drive improvements to cybersecurity standards and practices within the government. While many elements of this are already in place under the Federal Information Security Modernization Act (FISMA), no office or department is specifically tasked with measurement, root cause analysis, or incorporating those lessons into standards. These steps will improve the trajectory of America's cybersecurity, reduce our adversaries' ability to abuse our technology-enabled infrastructure, and reduce the costs of insecurity.

Below, we propose a variety of activities to be supported by different authorities or missions. Each is designed to create a stronger foundation for more precise evaluations in the future, addressing a long-standing challenge in cybersecurity in a new way. In contrast to other cybersecurity investments, the goal here is not the immediate treatment of a problem but the establishment of a system to assess whether problems are being treated effectively.

## *Focus on Whole-of-Government Efforts*

- **Include CPH Principles in the National Cybersecurity Strategy:** Call for explicitly including CPH principles and objectives into the National Cybersecurity Strategy to improve measurement and highlight gaps in available data.

- **Establish a Federal Task Force for CPH Coordination:** Create a task force comprising representatives from relevant federal agencies to coordinate CPH efforts across the government.

- **Develop a National CPH Framework:** Propose the creation of a comprehensive framework that outlines the principles, methodologies, and goals of CPH, incorporating measurement and specific outcome goals into cross-government initiatives.

## *Focus on Institutions and Data*

- **Establish a National Cybersecurity Public Health Institute:** Call for the creation of a dedicated federal institution to oversee CPH initiatives, similar to the CDC's role in public health. This institute would serve as a central hub for data gathering, analysis, and publication.

- **Create a National Cybersecurity Data Repository:** Propose the establishment of a centralized, secure database for collecting, analyzing, and sharing cybersecurity data across sectors. While narrower in scope than an institute, such a repository could still accelerate research efforts.

- **Implement Cybersecurity Health Reporting Standards:** Call for the development and implementation of standardized reporting mechanisms for cyber incidents and vulnerabilities, similar to [disease reporting in public health](). Consider whether these standards should be mandatory for certain professionals, specific sectors, or other criteria, and evaluate whether the current array of reporting standards can be simplified or made more useful for similar efforts.

- **Incorporate Long-Term Learning into Incident Notification and Reporting Systems:** Today's systems are often focused on immediate threat intelligence or consumer protection, and do not require information about attack mechanisms that enable learning. For many incidents, the attack mechanism is unknown. Knowing the fraction of incidents, and why they are unknown, would be a learning opportunity. A reporting system designed with long-term learning as a goal may prove more useful over time than current approaches focused solely on information sharing.

*Focus on Research*

- **Allocate Federal Funding for CPH Research:** Advocate for significant federal funding to support CPH research, including grants for academic institutions and private sector collaborations.

- **Develop Stress Tests for Cybersecurity Public Health in Critical Infrastructure:** Implement periodic stress tests to assess the resilience of critical infrastructure, supplemented by tabletops exercises or other simulations to identify systemic gaps or challenges.

*Focus on Creating a Community and Collaboration*

- **Create Incentives for Private Sector Participation:** Recommend the development of federal incentives, such as tax breaks or grants, to encourage organizations to actively participate in CPH initiatives.

- **Establish International CPH Partnerships:** Urge the government to take a leadership role in forming international partnerships and agreements aimed at advancing global cybersecurity public health.

*This quadrennial paper is part of a series compiled every four years by the **Computing Research Association (CRA)** and members of the computing research community to inform policymakers, community members, and the public about key research opportunities in areas of national priority. The selected topics reflect mutual interests across various subdisciplines within the computing research field. These papers explore potential research directions, challenges, and recommendations. The opinions expressed are those of the authors and CRA and do not represent the views of the organizations with which they are affiliated.*

*This material is based upon work supported by the U.S. National Science Foundation (NSF) under Grant No. 2300842. Any opinions, findings, and conclusions or recommendations expressed in this material are those of the authors and do not necessarily reflect the views of NSF.*